\documentclass[preprint]{aastex}

\usepackage{times}
\usepackage{wasysym} 
\usepackage{natbib} 

\newcommand{\degree}{\mbox{\ensuremath{^\circ}}}   %fractional degree symbol
\newcommand{\teff}{\mbox{$T_{\rm eff}$}}
\newcommand{\logg}{\mbox{$\log g$}}
\newcommand{\vsini}{\mbox{$v \sin i$}}
\newcommand{\mictrb}{\mbox{$v_{\rm mic}$}}
\newcommand{\mactrb}{\mbox{$v_{\rm mac}$}}

\newcommand{\kms}{\mbox{km\,s$^{-1}$}}
\newcommand{\halpha}{\mbox{H$_\alpha$}}

\newcommand{\rhostar}{\ensuremath{\rho_\star}}
\newcommand{\rhosun}{\ensuremath{\rho_\odot}}
\newcommand{\rhoj}{\ensuremath{\rho_{\rm J}}}
\newcommand{\rhopl}{\ensuremath{\rho_{\rm p}}}
\newcommand{\rj}{R\ensuremath{_{\rm J}}}
\newcommand{\mj}{M\ensuremath{_{\rm J}}}
\newcommand{\rsun}{R\ensuremath{_\odot}}
\newcommand{\msun}{M\ensuremath{_\odot}}
\newcommand{\rpl}{\ensuremath{R_{\rm p}}}
\newcommand{\mpl}{\ensuremath{M_{\rm p}}}
\newcommand{\rstar}{\ensuremath{R_\star}}
\newcommand{\mstar}{\ensuremath{M_\star}}

\shorttitle{WASP-37b}
\shortauthors{Simpson et al.}

\begin{document}

\title{WASP-37\lowercase{b}: a 1.8 M$_{\rm J}$ exoplanet transiting a metal-poor star}

\author{E. K. Simpson,$^1$ F. Faedi,$^1$ S. C. C. Barros,$^1$ D. J. A. Brown,$^2$ A. Collier Cameron,$^2$ 
L. Hebb,$^3$ D. Pollacco,$^1$ B. Smalley,$^4$ I. Todd,$^1$ O. W. Butters,$^5$ G. H\'{e}brard,$^6$  
J. McCormac,$^1$ G. R. M. Miller,$^2$ A. Santerne,$^7$, R. A. Street,$^8$ I. Skillen,$^9$ A. H. M. J. Triaud,$^{10}$  
D. R. Anderson,$^4$ J. Bento,$^{11}$ I. Boisse,$^6$ F. Bouchy,$^{6,12}$ B. Enoch,$^2$ C. A. Haswell,$^{13}$ 
C. Hellier,$^4$ S. Holmes,$^{13}$ K. Horne,$^2$ F. P. Keenan,$^1$ T. A. Lister,$^8$ P. F. L. Maxted,$^4$ V. Moulds,$^1$  C. Moutou,$^7$ A. J. Norton,$^{13}$ N. Parley,$^2$  F.  Pepe,$^{10}$ D. Queloz,$^{10}$ D. Segransan,$^{10}$ A. M. S. Smith,$^4$ 
H. C. Stempels,$^{14}$ S. Udry,$^{10}$ C. A. Watson,$^1$ R. G. West,$^5$ and P. J. Wheatley$^{11}$}

\affil{
$^1$ Astrophysics Research Centre, Queen's University Belfast, BT7 1NN, UK \\
$^2$ School of Physics and Astronomy, University of St Andrews, St Andrews, Fife KY16 9SS, UK\\
$^3$ Department of Physics and Astronomy, Vanderbilt University, Nashville, TN 37235, USA\\
$^4$ Astrophysics Group, Keele University, Staffordshire, ST5 5BG, UK\\
$^5$ Department of Physics and Astronomy, University of Leicester, Leicester, LE1 7RH \\
$^6$ Institut d'Astrophysique de Paris, UMR7095 CNRS, Universit\'e Pierre \& Marie Curie, France\\
$^7$ Laboratoire d'Astrophysique de Marseille, 38 rue Fr\'ed\'eric Joliot-Curie, 13388 Marseille, France \\
$^8$ Las Cumbres Observatory Global Telescope Network, 6740 Cortona Drive Suite 102, CA 93117, USA\\
$^9$ Isaac Newton Group of Telescopes, Apartado de Correos 321, E-38700 Santa Cruz de la Palma, Spain \\
$^{10}$ Observatoire de Gen\`eve, Universit\'e de Gen\`eve, 51 Ch. des Maillettes, 1290 Sauverny, Switzerland\\
$^{11}$ Department of Physics, University of Warwick, Coventry CV4 7AL, UK\\
$^{12}$ Observatoire de Haute-Provence, CNRS/OAMP, 04870 St Michel l'Observatoire, France\\
$^{13}$ Department of Physics and Astronomy, The Open University, Milton Keynes, MK7 6AA, UK\\
$^{14}$ Department of Physics and Astronomy, Uppsala University, Box 516, SE-751 20 Uppsala, Sweden\\
}

\begin{abstract}

We report on the discovery  of WASP-37b, a transiting hot Jupiter orbiting a $m_{\rm v}$ = 12.7 G2-type dwarf, with a period of 3.577469 $\pm$ 0.000011 d, transit epoch $T_{0}$ = 2455338.6188 $\pm$ 0.0006 (HJD)\footnote{Dates throughout the paper are given in Coordinated Universal Time (UTC)}, and a transit duration 0.1304$^{+ 0.0018}_{- 0.0017}$ d. The planetary companion has a mass \mpl\ = 1.80 $\pm$ 0.17 \mj\ and radius \rpl\ = 1.16$^{+ 0.07}_{- 0.06}$ \rj, yielding a mean density of 1.15$^{+ 0.12}_{- 0.15}$ \rhoj. From a spectral analysis, we find the host star has \mstar\ = 0.925 $\pm$ 0.120 \msun, \rstar\ = 1.003 $\pm$ 0.053 \rsun, \teff\ = 5800 $\pm$ 150 K and [Fe/H] = $-$0.40 $\pm$ 0.12. WASP-37 is therefore one of the lowest metallicity stars to host a transiting planet. 

\end{abstract}


\section{Introduction}\label{Intro}

Extrasolar planets show a huge diversity in their properties and this has important implications for theories of planet formation, structure and evolution. Systems with high orbital inclinations, in which the planet transits across the face of the host star as seen from Earth, are extremely valuable as they allow us to precisely measure many fundamental planetary properties, including radius, mass and density, which can be used to test these theories \citep{Haswell10}. 

The parameter space which we are able to explore with transiting planets is biased by instrumental and observational limitations. However, many challenges faced by the current surveys are being overcome by the ability to decrease systematic noise and optimise follow-up strategies. Although the majority of the $\sim$100 transiting planets thus far discovered are short period, Jupiter-sized objects, they show a remarkable variety in their physical and dynamical characteristics, such as the extreme eccentricity of HD 80606b \citep{Naef01,Laughlin09, Moutou09,Fossey09,Garcia09}, the ultra-short period of WASP-19b \citep{Hebb10} and the puzzlingly low densities of WASP-17b \citep{Anderson10} and Kepler-7b \citep{Latham10}. 

Here we describe the properties of a new transiting planet discovered by the SuperWASP survey, WASP-37b. SuperWASP has been a major contributor to the discovery of bright (9 $< m_{\rm v} <$ 13) transiting planets since it began operation in 2004 \citep{Pollacco06}. The project runs two stations, SuperWASP on La Palma, Canary Islands and WASP-S at SAAO in South Africa, each with a field of view of almost 500 square degrees. A number of recent upgrades have been implemented to reduce systematic noise and improve photometric precision. These include reducing temperature fluctuations, which cause changes in the camera focus, by installing heating tubes, air conditioning and improving dome insulation. As a consequence, the variation in the stellar FWHM during the course of a night has been halved. For more details, see \citet{Barros10}.

The planet host star WASP-37 resides in an equatorial region of the sky which is monitored by both WASP instruments, significantly increasing the amount of data collected on the target. It is accessible to observatories in both hemispheres and we present follow-up photometric and spectroscopic observations taken to establish the planetary nature of the transiting object and characterize it using the RISE (Liverpool Telescope), Spectral (Faulkes Telescope South), SOPHIE (1.93-m OHP) and CORALIE (Swiss 1.2-m) instruments.  

This paper is structured as follows: Section \ref{OandM} describes the observations, including the discovery data and photometric and spectroscopic follow-up. The results of the derived system parameters are presented in Section \ref{Results}, including the stellar and planetary properties. Finally we discuss our findings in Section \ref{Conc}.

\section{Observations} \label{OandM}
 
\subsection{SuperWASP photometry}

\begin{figure}[t]
\centering
\includegraphics[width=10cm]{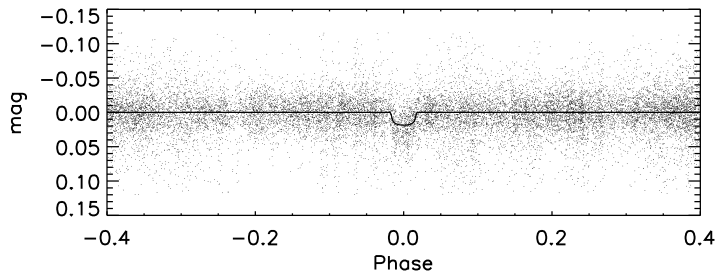}
\caption{The combined, unbinned SuperWASP light curve for WASP-37, folded on the orbital period of $P$ = 3.577 d. Superimposed is the model transit light curve, based on the system parameters determined from a global fit, see Section \ref{Planet}.  \label{SWASPlc} }
\end{figure} 

WASP-37 is a V = 12.7 mag star in the constellation Virgo, located at $\mathrm {\alpha_{J2000} = 14^{h}47^{m}46^{s}.57}$, \newline $\mathrm{\delta_{J2000} = +01\degr03\arcmin53\arcsec.9}$ (GSC 00326-00658; 2MASS 14474655+0103538). It was observed by SuperWASP (La Palma) between March and June in 2008 and 2009, and by WASP-S (South Africa) during June to July 2008 and March to July 2009. A total of 22,593 photometric data points were obtained during these intervals. The pipeline-processed data were de-trended and searched for transits using the methods described in \citet{Cameron06}. A periodic, transit-like signature was detected independently at the same period (3.58 d) in multiple cameras and in  successive seasons, strongly suggesting that the transits were real, despite the faint magnitude. The folded light curve is shown in Figure \ref{SWASPlc} with the best-fit model. 

The star underwent several consistency tests aimed at eliminating false positives. It is isolated within the WASP photometric aperture in the Digitized Sky Survey (DSS) image, ruling out resolved blends. We also checked for signatures of unresolved blends  \citep[see][]{Cameron07}: No significant ellipsoidal variability was measurable in the WASP light curve folded on the transit period, nor any signs of a secondary eclipse. Moreover, the transit depth and duration yielded a planet-like radius for the companion, and a stellar density appropriate to a main-sequence host star of the effective temperature derived from the 2MASS colours. Having passed all these tests, the star was selected for follow-up observations.

\subsection{Photometric follow-up}

\begin{figure}[t]
\centering
\includegraphics[width=8.5cm]{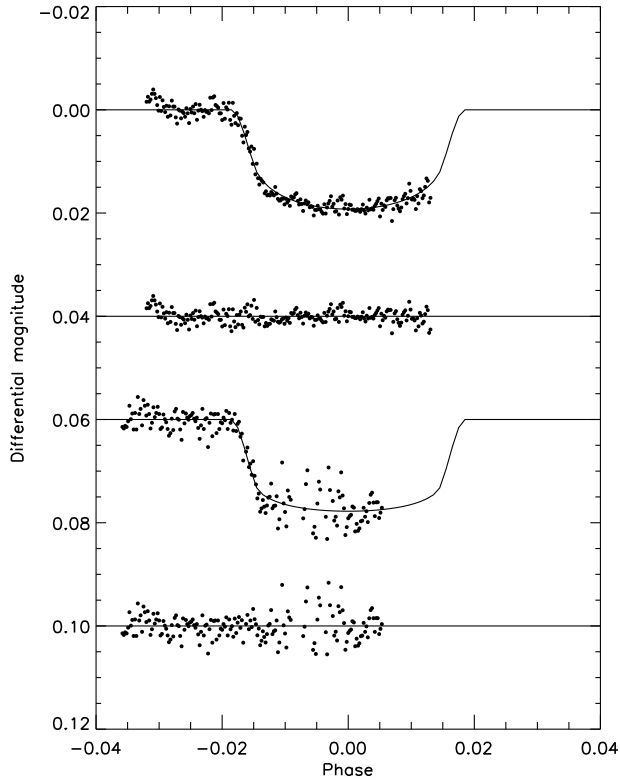}
\caption{Photometry of transit events of WASP-37b on 2010 May 21 and 2010 June 30. The light curves have been offset from zero by arbitrary amounts for clarity. Superimposed are the model transit light curves based on the determined system parameters, see Section \ref{Planet}. Residuals from the fit are displayed underneath. \label{lcs} }
\end{figure}
  
We obtained two further transit light curves of WASP-37 in order to refine the photometric parameters, and they are shown in Figure \ref{lcs}. All photometric data presented in this paper are available from the NStED database.\footnote{http://nsted.ipac.caltech.edu}

A partial transit was observed with LT/RISE on 2010 May 21. The full transit was not obtainable as the star set below the observing limits of the LT prior to egress. RISE is a frame transfer CCD located on the robotic 2.0-m Liverpool Telescope (LT) on La Palma with a broad band V + R filter \citep{Steele08, Gibson08}. The telescope was defocused by $-$1.0 mm to give FWHM = 17 pixels = 9.2 arcsec, and the CCD was used in 2$\times$2 binning mode with an exposure time of 65 s and effectively no dead time. This allowed 215 images to be taken over the 3.9 h period, including 1 h of observations before transit. The data were reduced using the ULTRACAM pipeline \citep{Dhillon07} and differential photometry was performed relative to five nearby bright stars using an 18 pixel radius aperture. 

A further partial transit was obtained on 2010 June 30 with the Spectral\footnote{http://lcogt.net/en/network/2m} camera (FS01) on the LCOGT 2.0-m Faulkes Telescope South (FTS, Siding Spring, Australia). The Pan-STARRS-Z filter was used with the instrument in binning 2$\times$2 mode, giving 0.303 arcsec/pixel, and no defocussing. The exposure time was 45 s, and 169 images were taken in the 3.5 h period. Data were reduced using a pipeline\footnote{http://telescope.livjm.ac.uk/Info/TelInst/Pipelines} written at Liverpool John Moores University, then differential photometry performed using the IRAF/DAOPHOT\footnote{IRAF is distributed by National Optical Astronomy Observatories, operated by the Association of Universities for Research in Astronomy, Inc., under contract with the National Science Foundation, USA.} package. Eight comparison stars were used with an 8 pixel radius aperture. Points with error bars larger than 0.01 mag were removed from the second half of the dataset. These were likely caused by passing cloud, and the observation was stopped prematurely due to bad weather.

\subsection{Spectroscopic follow-up }

\begin{deluxetable}{lccrc}
\tablewidth{0pc}
\tabletypesize{\scriptsize}
\tablecaption{Radial velocity (RV) and line bisector span ($V_{\rm span}$ ) measurements of WASP-37. \label{rv-data} } 
\tablehead{
	\colhead{BJD} & 
	\colhead{RV} & 
	\colhead{\ensuremath{\sigma_{\rm RV}}} & 
	\colhead{$V_{\rm span}$} & 
	\colhead{Instrument} \\	
	\colhead{\hbox{(2,400,000+)}} & 
	\colhead{(\kms)} & 
	\colhead{(\kms)} &
	\colhead{(\kms)} &
	\colhead{}
}
\startdata
$55304.4975$ &	$7.892$ &$	0.012$ &	 $0.024$ & SOPHIE\\
$55305.4660 $&	$8.223$ &$	0.014 $&	$-0.028$ & SOPHIE\\
$55334.5449 $&	$8.196$ &	$0.019 $&	 $0.004$ & SOPHIE\\
$55336.4764 $&	$7.825$ &	$0.021 $&	$-0.088$ & SOPHIE\\
$55345.4618 $&	$8.115$ &	$0.018 $&	 $0.013$ & SOPHIE\\
$55305.7532  $&	 $8.152$  &$   0.018 $ &	$-0.024$   & CORALIE \\	
$55310.8236  $&	  $7.678$  &$   0.043 $&	$-0.029 $  &CORALIE\\		
$55311.8514  $&	  $7.919$ &  $ 0.025  $&	$-0.033 $  &CORALIE\\		
$55312.8266  $&	  $8.064$  &  $ 0.031 $ &	$-0.091$   &CORALIE\\		
$55321.7716  $&	  $7.668$  &  $ 0.023 $ &	$0.058 $   & CORALIE\\		
$55324.7114  $&	  $7.769$  &  $ 0.022 $ &	$-0.011$   &CORALIE\\		
$55325.6868  $&	  $7.776$  &   $0.024 $ &	$-0.025$   &CORALIE\\		
$55376.6684  $&	  $8.103$  &   $0.031 $ &	$-0.004$   &CORALIE\\
\enddata
\end{deluxetable}

In order to establish the planetary nature and determine the stellar parameters, we obtained follow-up spectroscopic observations. Five spectra were taken  between 2010 April 17 and 2010 May 28 with the stabilised echelle spectrograph SOPHIE at the 1.93-m telescope of Observatoire de Haute-Provence \citep{Perruchot08,Bouchy09}. The observations were all made with a signal-to-noise ratio of S/N$\sim$20 in order to minimise the Charge Transfer Inefficiency (CTI) effect \citep{Bouchy09}. Two 3 arc-second diameter optical fibres were used, the first centred on the target and the second on the sky to simultaneously measure the background to remove contamination from scattered moonlight. A further 8 spectra were obtained with the CORALIE Fibre-Fed Echelle Spectrograph on the Swiss 1.2-m telescope at ESO-La Silla, Chile between 2010 April 19 and 2010 June 29 with S/N$\sim$10--20 in dark/grey time to minimise contamination from scattered moonlight. The data were processed using the standard pipeline \citep{Baranne96,Pepe02}.  

\begin{figure}[t] 
\centering
\includegraphics[width=10cm]{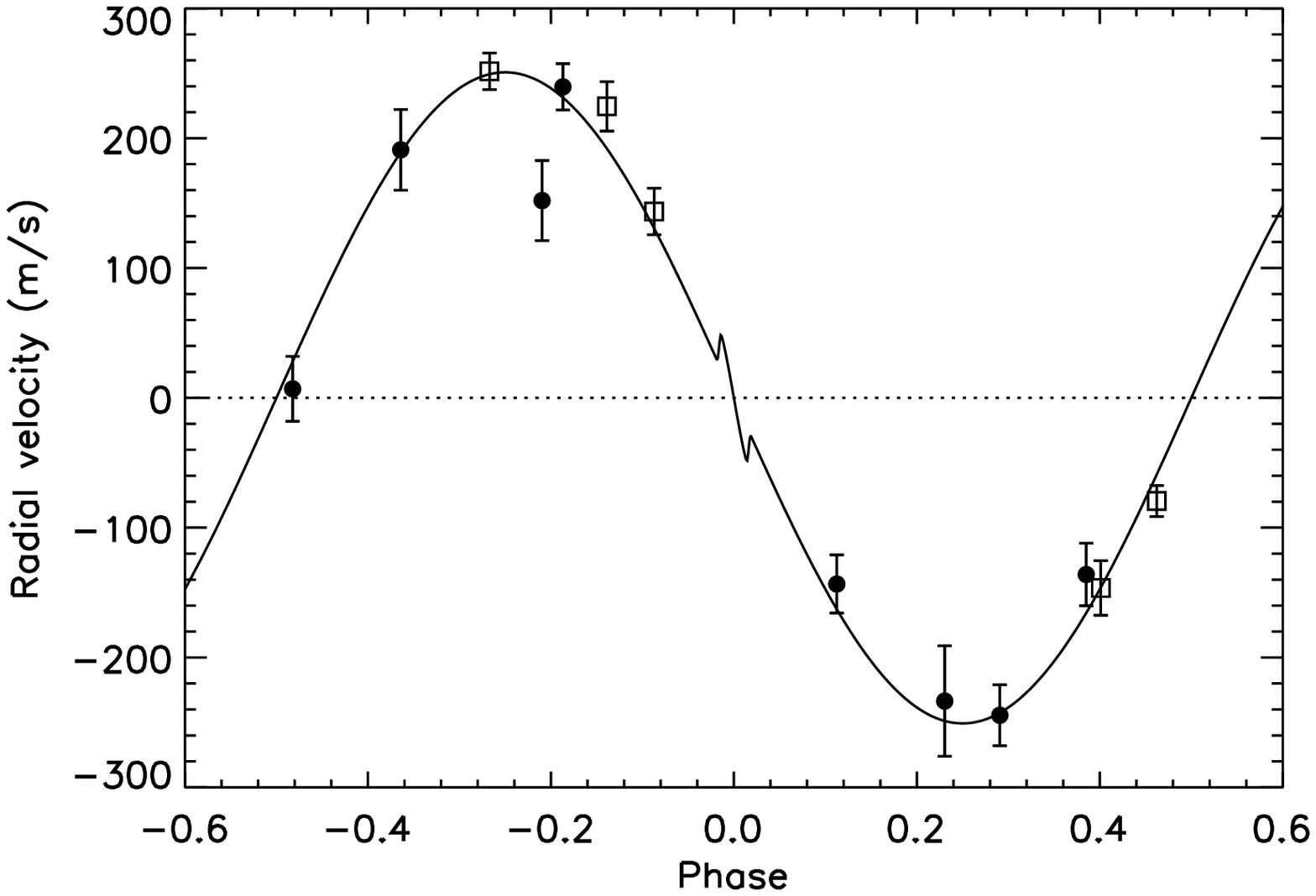}
\includegraphics[width=10cm]{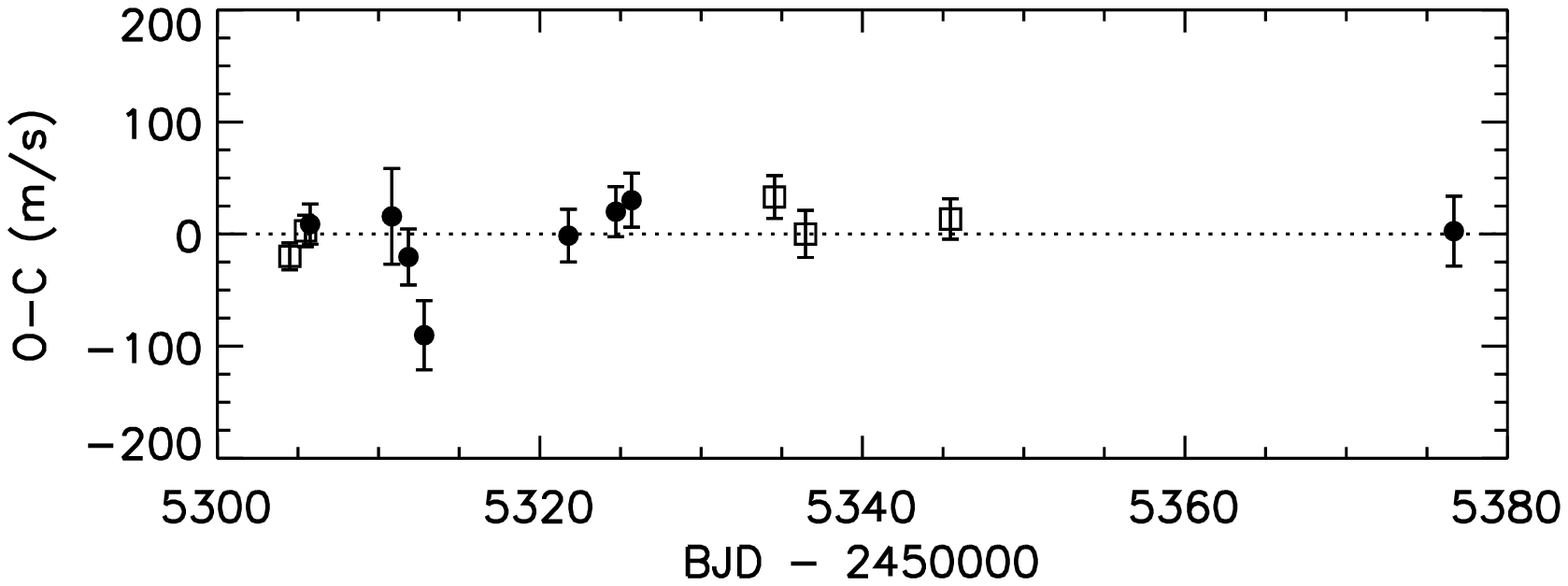}
\includegraphics[width=10cm]{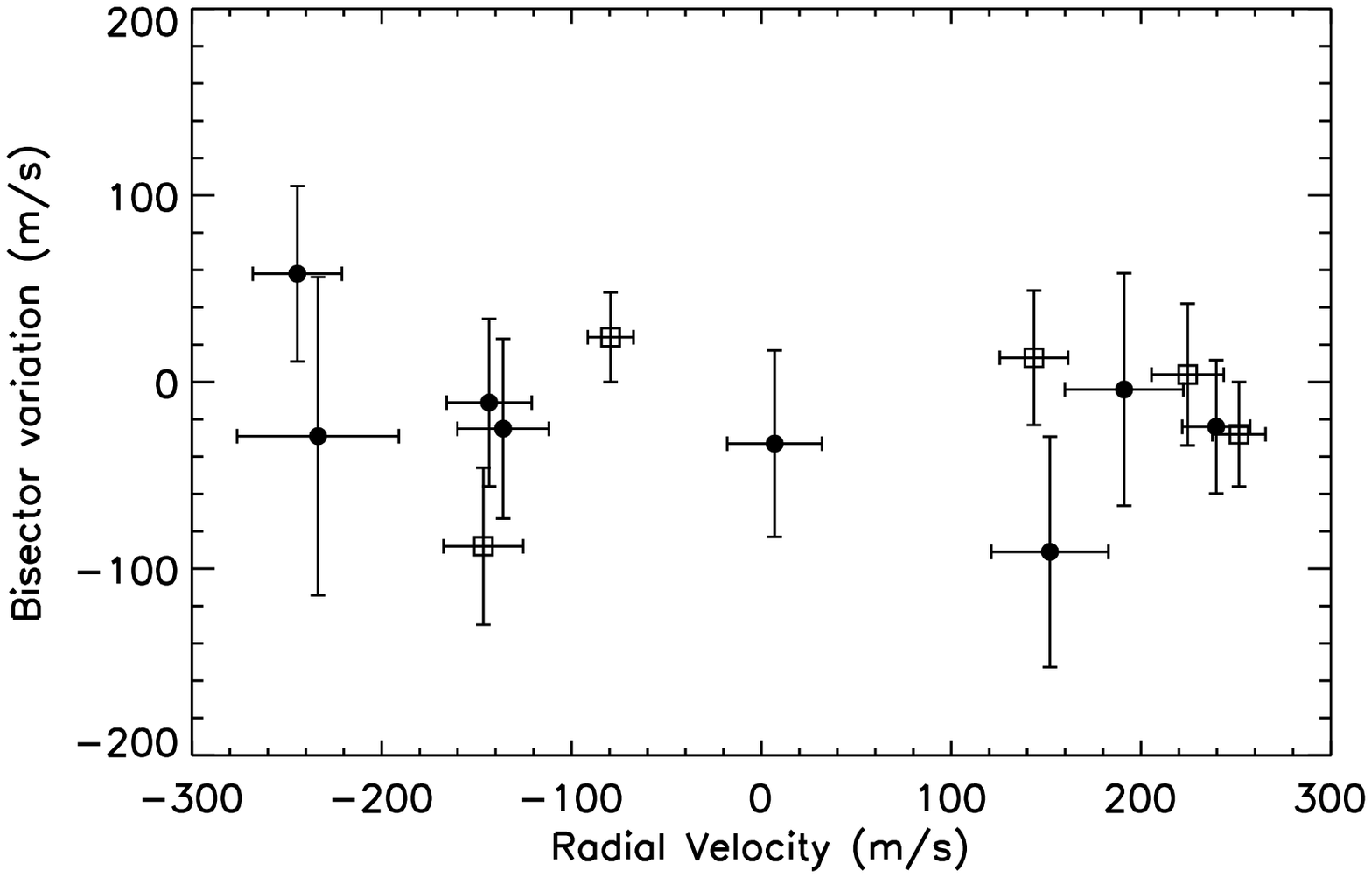}
\caption{\textit{Upper panel:} Phase folded radial velocity measurements of WASP-37, combining data from the Swiss 1.2-m/CORALIE (filled circles) and OHP/SOPHIE (open squares), and superimposed with the best-fit model RV curve based on the determined system parameters, see Section \ref{Planet}. The centre-of-mass velocities, $\gamma_{\rm SOPHIE}$ = 7.9714 \kms\ and
$\gamma_{\rm CORALIE}$ = 7.9116 \kms\ were subtracted from the RVs. The predicted Rossiter-McLaughlin effect is shown for a spin orbit aligned system with \vsini\ = 2.4 $\pm$ 1.6 \kms. \textit{Middle panel:} Residuals from the orbital fit plotted against time. No long-term trend is visible. \textit{Lower panel:} The bisector span measurements as a function of radial velocity, showing no correlation. The uncertainties in the bisectors were taken as twice the RV uncertainties.} 
\label{RV} 
\end{figure}

The radial velocities (RVs) and line bisector spans \citep*[$V_{\rm span}$, see][]{Toner88} derived from cross correlation are shown in Table \ref{rv-data} and plotted with the best-fit model in Figure \ref{RV}. No significant correlation is seen between the bisector span and radial velocity, with a supporting the signal's origin as a planetary companion rather than a blended eclipsing binary system \citep{Queloz01}. A Spearman rank-order correlation test also indicates that any correlation between the radial velocities and bisector is weak, with a probability of correlation of 0.12.

\section{Results} \label{Results}

\begin{deluxetable}{lc}
\tablewidth{0pc}
\tabletypesize{\scriptsize}
\tablecaption{Stellar parameters of WASP-37 \label{sparams}} 
\tablehead{
	\colhead{Parameter (Unit)} & 
	\colhead{Value}
}
\startdata
\sidehead{Photometric \& spatial properties:}
RA (J2000) & 14:47:46.57\\
DEC (J2000) & +01:03:53.9 \\
$V$ (mag) &$12.704 \pm 0.149$ \\
$B-V$ (mag) & $0.600 \pm 0.050$ \\ 
$J$ (mag)& $11.499 \pm 0.022$ \\
$H$ (mag)& $11.181 \pm 0.026$ \\
$K_{\rm s}$ (mag)& $11.093 \pm 0.023$\\
$\mu_{\rm RA}$ (mas year$^{-1}$) & $-23.2 \pm 5.4$\\
$\mu_{\rm DEC}$ (mas year$^{-1}$) & $22.8 \pm 5.4 $\\
$U$ (\kms) & $-27.2 \pm 3.4$  \\
$V$ (\kms) & $7.8 \pm 14.0$  \\
$W$ (\kms) & $45.5 \pm 3.4$  \\
Galactic  longitude ($l$)  (deg) & $355$  \\
Galactic latitude ($b$) (deg) & $52$\\
\sidehead{Spectroscopic properties:}
\teff\ (K)     & $5800 \pm 150$ \\
\logg\ (cgs)     & $4.25 \pm 0.15$ \\
\mictrb\ (\kms)   & $1.0 \pm 0.2$ \\
\mactrb\ (\kms) &  $3.6 \pm 0.3$ \\
\vsini\ (\kms)  & $2.4 \pm 1.6$ \\
{[Fe/H]}   &$-0.40 \pm 0.12$ \\
{[Na/H]}   &$-0.38 \pm 0.16$ \\
{[Mg/H]}   &$-0.12 \pm 0.10$ \\
{[Si/H]}   &$-0.27 \pm 0.08$ \\
{[Ca/H]}   &$-0.25 \pm 0.13$ \\
{[Sc/H]}   &$-0.19 \pm 0.16$ \\
{[Ti/H]}   &$-0.28 \pm 0.14$ \\
{[V/H]}    &$-0.30 \pm 0.15$ \\
{[Cr/H]}   &$-0.37 \pm 0.16$ \\
{[Mn/H]}   &$-0.56 \pm 0.10$ \\
{[Co/H]}   &$-0.41 \pm 0.17$ \\
{[Ni/H]}   &$-0.37 \pm 0.08$ \\
log A(Li)  &   $2.23 \pm 0.13$ \\
$E(B-V)$ &$0.05$ \\
\sidehead{Derived properties:}
\mstar\ (\msun) & $0.925 \pm 0.120$   \\
\rstar\ (\rsun)&  $1.003 \pm 0.053$  \\
\rhostar\ (\rhosun) & $ 0.931^{+0.064}_{-0.099}$ \\
$L_{*}$ (L$_{\odot}$) &  $0.953_{-0.114}^{+0.184}$ \\
Age (Gyr) & $11^{+3}_{-4}$ \\
Distance (pc) & $343 \pm 36$ \\
Spectral type & G2V \\
\enddata
\tablecomments{The photometric and spatial properties are taken from or derived using the following sources: $V$ \citep[TASS,][]{TASS}, $B-V$ \citep{Casagrande10}, $J, H, K_{\rm s}$ \citep[2MASS,][]{2MASS}, proper motions \citep[NOMAD,][]{NOMAD}, space velocities (see Section \ref{check}) and galactic co-ordinates (NED). The spectroscopic and derived properties are determined from the spectroscopic analysis described in Section \ref{stellar}. \mstar\ is the mean of the masses found from the empirical relationship of \citet{Torres10} and stellar models; \rhostar\ is found from the light curve geometry; and the stellar radius is derived from \rhostar\ and \mstar. $L_{*}$ and age are determined from stellar models.} \\
\end{deluxetable}

\subsection{Stellar parameters}\label{stellar}

We used several techniques to infer the fundamental stellar properties of WASP-37 and the results are shown in Table \ref{sparams}. We first performed a spectral analysis using the methods given in \citet{Gillon09b}. The individual CORALIE and SOPHIE spectra from the standard reduction pipeline were co-added to produce a single spectrum with an average S/N of around 60:1. The \halpha\ line was used to determine the effective temperature (\teff), while the Na {\sc i} D and Mg {\sc i} b lines were used as surface gravity (\logg) diagnostics. This yielded the values \teff\ = 5800 $\pm$ 150 K and \logg\ = 4.25 $\pm$ 0.15. The effective temperature indicates that the star is of spectral type G2 \citep{Gray08}. 

The value for microturbulence (\mictrb) was determined from Fe~{\sc i} using the \citet{Magain84} method. We assumed a value for macroturbulence (\mactrb) of 3.6 $\pm$ 0.3 \kms\, based on the tabulation by \citet{Gray08}, and an instrumental FWHM of 0.11 $\pm$ 0.01\AA, determined from the telluric lines around 6300\AA. The projected stellar rotation velocity (\vsini) was determined by fitting the profiles of several unblended Fe~{\sc i} lines, giving a best-fitting value of \vsini\ = 2.4 $\pm$ 1.6 \kms. 

Elemental abundances were determined from equivalent-width measurements of several clean and unblended lines. The quoted errors include the uncertainties in \teff, \logg\ and \mictrb, as well as the scatter due to measurement and atomic data uncertainties. We obtain [Fe/H] = $-$0.4 $\pm$ 0.12 and log A(Li) = 2.23 $\pm$ 0.13. Interstellar Na D lines are present in the spectra with equivalent widths of $\sim$100--150 m\AA, indicating an extinction of $E(B-V)$ = 0.05 using the calibrations of \citet*{Munari97}. 

\begin{figure}[t]
\centering
\includegraphics[width=8.5cm, angle=-90]{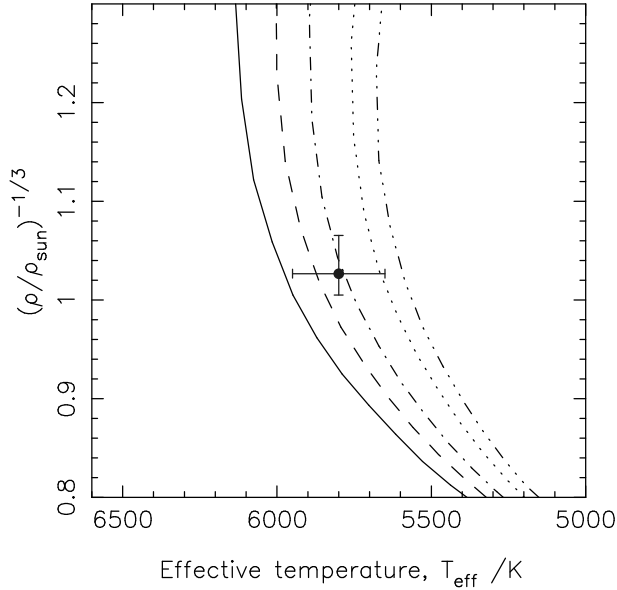}
\caption{Theoretical stellar isochrones from \citet{Yi01} plotted as the density proxy $R_{*}/M_{*}^{1/3}$ against effective temperature for [Fe/H] = $-$0.4 and ages 7, 9, 11, 14 and 16 Gyr (from left to right). The position of WASP-37 is indicated within the error ranges. }
\label{iso}
\end{figure}

Transit light curves provide a measure of the mean stellar density (\rhostar) which can be used as a luminosity indicator for stellar evolutionary models  \citep{Sozzetti07}. We used a Markov Chain Monte Carlo (MCMC) approach to globally model the photometric and radial velocity data (see Section \ref{Planet}) and obtained \rhostar\ = 0.931$^{+0.064}_{-0.099}$ $\rho_{\odot}$. Values of \teff, [Fe/H] and \rhostar\ were compared with the theoretical stellar evolutionary models of \citet{Yi01} to obtain the following stellar properties: \mstar\ = 0.849$^{+0.067}_{-0.040}$ \msun, \rstar\ = 0.977$^{+0.045}_{-0.042}$ \rsun, $L_{*}$ = 0.953$_{-0.114}^{+0.184}$ L$_{\odot}$, \logg$_{\rm iso}$ = 4.39$^{+0.02}_{-0.03}$ and an age of 11$^{+3}_{-4}$ Gyr. The model isochrones are shown in Figure \ref{iso}.

We find that the measurement of \logg\ from evolutionary models, \logg$_{\rm iso}$ = 4.39$^{+0.02}_{-0.03}$, is consistent with that found from the spectroscopic analysis, \logg\ = 4.25 $\pm$ 0.15. We re-fitted the isochrones using the 1$\sigma$ upper and lower limits of [Fe/H] and found that this affected the age and mass determinations somewhat, increasing the uncertainties in the age (8--13 Gyr) and stellar mass (0.806--0.914 \msun).

We computed the stellar distance using the following two methods. Firstly, the absolute bolometric magnitude, (calculated from the stellar luminosity) was compared to the extinction corrected V-band magnitude from TASS \citep[$m_{\rm v}$ = 12.7,][]{TASS}. The bolometric correction was taken as the range of values for a G2-type star given in \citet{DL} and \citet{Gray08}. This yielded a distance $D$ = 338 $\pm$ 41 pc. Secondly, we compared the angular diameter $\Theta$ = $2$ \rstar\ / $D$ = 0.025 $\pm$ 0.005 mas derived from the flux fitting method (see Section \ref{check})  to the stellar radius, yielding a distance $D$ = 360 $\pm$ 75 pc. It is reassuring that these values are similar, given that both methods rely on parameters from the stellar models and are therefore not independent. We take the distance to be the weighted mean of the two values, $D$ = 343 $\pm$ 36 pc.

\subsubsection{Independent checks}\label{check}

The stellar mass can also be determined through an empirical calibration between \teff, \rhostar\ and [Fe/H] found from eclipsing binaries \citep{Torres10, Enoch10}. This yields a larger mass, \mstar\ = 1.002 $\pm$ 0.034 \msun\ than that from the evolutionary model fit,  \mstar\ = 0.849$^{+0.067}_{-0.040}$ \msun. A possible cause is that WASP-37 is on the edge of the calibration where there are no stars of similar \teff\ and [Fe/H] and \logg. A similar situation was found for WASP-21, see \citet{Bouchy09}. In order to obtain a robust result, we have taken the stellar mass to be the average of the two values \mstar = 0.925 $\pm$ 0.120 \msun. 

We used archival magnitudes from 2MASS, DENIS, NOMAD, TASS and CMC14 to derive an independent value of the stellar temperature using the flux-fitting method \citep{Morossi85}. We find \teff$_{\rm flux}$ = 5840 $\pm$ 140 K, in excellent agreement with the temperature derived from spectral analysis.

Independent measures of \vsini\ and [Fe/H] can be derived from the shape of the spectral cross correlation function (CCF). Using the empirical calibrations  of \citet{Boisse10} for the SOPHIE HE mode, and inputing Contrast = 4.4 $\pm$ 1.3\% and FWHM = 9.48 $\pm$ 0.06 \kms\ from the SOPHIE CCF and $B-V$ = 0.60 $\pm$ 0.05 \citep[calculated using][]{Casagrande10}, we find [Fe/H]$_{\rm CCF}$ = $-$0.32 $\pm$ 0.17 and \vsini$_{\rm CCF}$ =  3.2 $\pm$ 1.0 \kms. These are also in excellent agreement with the values found from spectral analysis. 

The stellar evolution modelling provides colour indices which can be compared to observational values. We corrected the 2MASS magnitudes for extinction using $A(V)$ = 3.1 $\times E(B-V)$, appropriately scaled for each passband following \citet*{Cardelli89}. The model colours were transformed from the ESO to the 2MASS system following \citet{Carpenter01}. The 2MASS and model colours agree well, as shown in Table \ref{colors}.

\begin{deluxetable}{lccc}
\tablewidth{0pc}
\tabletypesize{\scriptsize}
\tablecaption{Derived colors of WASP-37  \label{colors} } 
\tablehead{
	\colhead{} & 
	\colhead{$J-K$} &
	\colhead{$J-H$}&
	\colhead{$H-K$}
}
\startdata
Model & $0.377 \pm 0.026$ & $0.293 \pm 0.026$ & $0.019 \pm 0.010$ \\
2MASS & $0.380 \pm 0.032$ & $0.304 \pm 0.034$ & $0.076 \pm 0.035$ \\
\enddata
\tablecomments{The model colors are derived from stellar evolutionary models and transformed to the 2MASS system. The 2MASS colours are corrected for extinction. See Section \ref{check} for details.}
\end{deluxetable}

The large relative error on \vsini\ provides little constraint on the maximum stellar rotation period  $P_{\rm rot} < 2 \pi R_{*} / v \sin i < 21^{+44}_{-9}$ d, or gyrochronological age; 1--28 Gyr \citep{Barnes07}. Equally, the lithium abundance only implies a lower limit on the age of 1--2 Gyr \citep*{SestitoRandich05}. 

Given the low metallicity of WASP-37, we investigated whether it could be a member of the thick disk using the method outlined in \citet{Bouchy10}. We calculated the space velocities according to \citet{Johnson87}, using the proper motions, distance and $\gamma$ given in Tables \ref{sparams} and \ref{pparams}. The velocities were updated to J2000 and corrected for solar motion using data from \citet{Dehnen98} giving $U$ = $-$27.2 $\pm$ 3.4 \kms, $V$ = 7.8 $\pm$ 14.0 \kms\ and $W$ = 45.5 $\pm$ 4.4 \kms\ (right-hand coordinate system).

In order to quantify the probability that WASP-37 is a member of the thick disk, we compared its metallicity and space velocities to the properties of similar stars in the Besan\c{c}on Galactic model \citep{Robin03}. We generated 78,000 model stars in 20 realisations and returned all stars with apparent magnitudes between $m_{\rm v}$ = 8--16, spectral type F5--K9 and luminosity class IV and V for 30 square degrees in the direction of WASP-37 (galactic coordinates $l$ = 356\degree, $b$ = 53\degree, NED). For the stars with [Fe/H] and space velocities within 3$\sigma$ of WASP-37, half were members of the old thin disk ($>$6 Gyr), one third were part of the thick disk, and the remainder were 1.5--6 Gyr thin disk members. Abundance analysis shows that WASP-37 has enhanced alpha elements and although this is not accounted for in the simulation, we estimate that it increases the probability of thick disk membership but that it is also equally likely to be part of the old thin disk. It is reassuring that this independent analysis also finds that WASP-37 is part of an old population.

\begin{deluxetable}{lc}
\tablewidth{0pc}
\tabletypesize{\scriptsize}
\tablecaption{System parameters of WASP-37b \label{pparams} } 
\tablehead{
	\colhead{Parameter (Unit)} & 
	\colhead{Value}
}
\startdata
\sidehead{Photometric parameters:}
$P$ (d) & $3.577469 \pm 0.000011 $\\
$T_{0}$ (HJD) & $2455338.6188 \pm 0.0006$\\
$T_{\rm dur}$ (d) & $0.1304^{+ 0.0018}_{- 0.0017}$\\
$\Delta F = (R_{\rm P}/R_{*})^{2}$ & $0.01427^{+ 0.00030}_{- 0.00023}$\\
$R_{\rm P}/R_{*}$ &  0.11946$^{+ 0.00126}_{- 0.00096}$\ \\
$a/R_{*}$ & $9.567 \pm 0.648$  \\
$b$ & $0.198^{+ 0.132}_{- 0.128}$\\
$i (^\circ)$ & $88.82^{+ 0.77}_{- 0.86}$\\
\sidehead{Spectroscopic parameters:}
$K$ (\kms) & $0.2507 \pm 0.0084$\\
$\gamma_{\rm SOPHIE}$ (\kms) & $7.9714 \pm 0.0024$\\
$\gamma_{\rm CORALIE}$ (\kms) & $7.9116 \pm 0.0024$ \\
$e$ & $0$ (fixed)\\
\sidehead{Derived parameters:}
\mpl\ (\mj) & $1.80 \pm 0.17$\\
\rpl\ (\rj) & $1.16^{+ 0.07}_{- 0.06}$ \\
$\rho_{\rm P} (\rho_{\rm J})$ & $1.15^{+ 0.12}_{- 0.15}$\\
$\log g_{\rm P}$ (cgs) & $3.48^{+ 0.03}_{- 0.04}$\\
$a$ (AU)  & $0.0446 \pm 0.0019$\\
$T_{\rm eq, A=0}$ (K) & $1323^{+ 25}_{- 15}$\\
\enddata
\end{deluxetable}

\subsection{Planet parameters}\label{Planet}

\begin{deluxetable}{lccccc}
\tablewidth{0pc}
\tabletypesize{\scriptsize}
\tablecaption{Limb darkening coefficients \label{ld} } 
\tablehead{
	\colhead{Light curve} &
	\colhead{Band} & 
	\colhead{$a_{1}$} &
	\colhead{$a_{2}$}&
	\colhead{$a_{3}$}&
	\colhead{$a_{4}$}
}
\startdata
SuperWASP & R & 0.643 & -0.278  & 0.800 & -0.413 \\
RISE & V & 0.570 & -0.131 & 0.791 & -0.420  \\
FTN & Z & 0.623 & -0.331 &0.672 & -0.346 \\
\enddata
\end{deluxetable}

To determine the properties of the planet, we simultaneously modelled the SuperWASP, LT and FTN light curves and the CORALIE and SOPHIE radial velocities with a global MCMC fit.  Details of this process are described in \citet{Cameron07} and \citet{Pollacco08}. The free parameters in the fit are: orbital period $P$; transit epoch $T_{0}$; transit duration $T_{\rm dur}$;  ratio of planet to star radius $(R_{\rm p}/R_{*})^{2}$; impact parameter $b$; RV semi-amplitude $K$; Lagrangian elements $e\cos \omega$ and $e \sin \omega$ where $e$ is the eccentricity and $\omega$ is the longitude of periastron; and the systematic offset velocity $\gamma$. In this particular case, two systematic velocities were fit to allow for instrumental offsets between the SOPHIE and CORALIE datasets.

We propagated the uncertainty in the stellar mass through to the derived parameters (\mstar\ = 0.925 $\pm$ 0.120 \msun, see Section \ref{check}).  We used the 4 coefficient \citet{claret00, claret04} non-linear limb darkening coefficients for the appropriate stellar temperature and photometric passband appropriate for each light curve, see Table \ref{ld}. The limb darkening coefficients are re-calculated at each step in the MCMC chain to take into account the uncertainties in the stellar temperature and radius. 

The photometric error bars were scaled to account for any underestimations so that the best fitting model results in $\chi^{2}_{\rm red}$ = $\chi^{2}$/dof = 1 (dof = number of points $-$ number of fitted parameters). The radial velocity error bars did not require rescaling with a jitter term in order to obtain $\chi^{2}_{\rm red}$ = 1. The RMS of the residuals from the best-fit model are: RMS$_{\rm WASP}$ = 0.0292 mag,  RMS$_{\rm LT}$ = 0.0013 mag,  RMS$_{\rm FTS}$ = 0.0025 mag,  RMS$_{\rm SOPHIE}$ = 0.0194 \kms and RMS$_{\rm CORALIE}$ = 0.0380 \kms. By removing the outlying point at 8.064 \kms, we obtain RMS$_{\rm CORALIE}$ = 0.0164 \kms.

An initial fit was performed in which the eccentricity was allowed to float, yielding $e$ = 0.036 $\pm$ 0.021. From the Lucy-Sweeney test \citep{Lucy71}, we find it is only significant at the 1$\sigma$ level, hence a circular orbit was adopted ($\chi^{2}_{\rm circ}$ = 18.3, $\chi^{2}_{\rm ecc}$ = 14.0 with 13 RVs and 5 free parameters: $K$, $\gamma_{\rm CORALIE}$, $\gamma_{\rm SOPHIE}$, $e\cos \omega$ and $e \sin \omega$). We determined an upper limit for the eccentricity as $e_{\rm upper} = e + 3\sigma = 0.078$. We also fitted a linear trend in the RVs to search for a third body in the system, but found that the decrease in $\chi^{2}$ did not warrant the extra free parameter. From the F-test, we found the significance of the trend was less than 1$\sigma$ ($\chi^{2}$ = 18.3, $\chi^{2}_{\rm trend}$ = 15.4 with 13 RVs and 4 free parameters: $K$, $\gamma_{\rm CORALIE}$, $\gamma_{\rm SOPHIE}$ and linear trend $\dot{\gamma}$). 

We performed the final MCMC fit with a chain length of 20,000 points and the resulting best-fit parameters and uncertainties are shown in Table \ref{pparams}. WASP-37b has $P$ = 3.577469 $\pm$ 0.000011 d, \mpl\ = 1.80 $\pm$ 0.17 \mj, \rpl\ =  1.16$^{+ 0.07}_{- 0.06}$  \rj\ and \rhopl\ = 1.15$^{+ 0.12}_{- 0.15}$ \rhoj.

\section{Discussion} \label{Conc}

\begin{figure}[t]
\centering
\includegraphics[width=10cm]{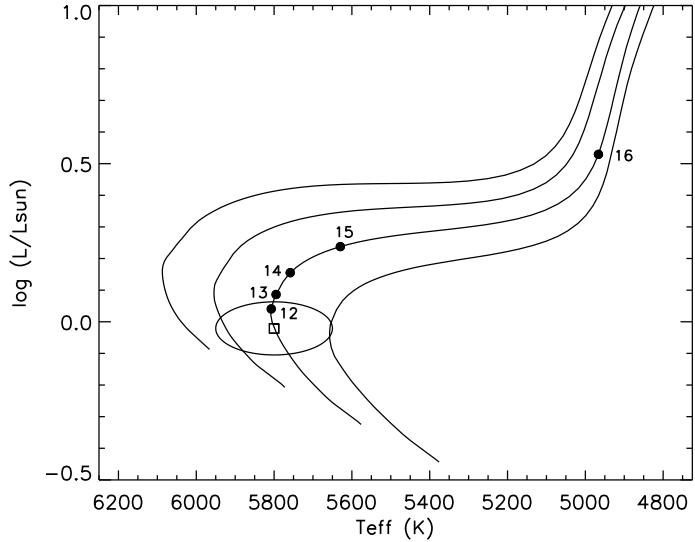}
\caption{Evolutionary mass tracks from \citet{Yi01} for [Fe/H] = $-$0.4 and stellar masses of 0.95, 0.9, 0.85 and 0.8 \msun\ (from left to right). The current position of WASP-37 is shown as a square within the 1$\sigma$ uncertainty ellipse. The 12, 13, 14, 15 and 16 Gyr points are shown on the best-fit, 0.85 \msun\ track. WASP-37 is predicted to evolve onto the subgiant branch at age $\sim$15 Gyr and red giant branch $\sim$1 Gyr later.}
\label{evo}
\end{figure}

We report the discovery of a new transiting hot Jupiter, WASP-37b. The planet has a radius comparable to Jupiter (\rpl\ = 1.16$^{+ 0.07}_{- 0.06}$ \rj) but is more massive (\mpl\ = 1.80 $\pm$ 0.17 \mj) and hence is very similar to WASP-5 \citep{Anderson08, Southworth09}. The surface gravity of WASP-37b is high compared to other planets with similar orbital periods (\logg$_{\rm p}$ = 3.48$^{+ 0.03}_{- 0.04}$, $g_{\rm p}$ = 30.2$^{+2.0}_{-2.8}$ ms$^{-2}$, \rhopl\ = 1.15$^{+ 0.12}_{- 0.15}$ \rhoj), which lies significantly above the period-gravity correlation proposed by \citet{Southworth10}. 

By comparison with the theoretical models of \citet*{Fortney07} and \citet*{Baraffe08}, we find that WASP-37b has an inflated radius. The models do not cover the age range of WASP-37, however given that planetary radii are thought to decrease with age, the expected radius of WASP-37b will be smaller than those given for the oldest model. For an equivalent semi-major axis with solar insolation of 0.445 AU, the measured mass and radius do not suggest the presence of a core of heavy elements for models older than 1 Gyr, and this is consistent with the correlation between core mass and metallicity \citep{Guillot06, Burrows07} given the low stellar metallicity.

We detect no significant orbital eccentricity and find an upper limit of 0.098. Nor do we find evidence of any long-term trend caused by a third body. Our radial velocity observations span $\sim$ 70 days and continued monitoring would reveal the presence of any longer-term trends and further constrain the eccentricity.

With a stellar effective temperature of 5800 $\pm$ 150 K and orbital period of 3.577469 $\pm$ 0.000011 d, the planet is predicted to have a equilibrium temperature of 1323$^{+25}_{-15}$ K (assuming zero albedo and full redistribution of the incident radiation). Secondary eclipse measurements of WASP-37 may be a way of constraining the influence of metallicity on the presence or not of atmospheric temperature inversions.  

The relatively large uncertainties in the stellar parameters are caused  by the low S/N of the combined spectrum used for spectral analysis, due to the faintness of the target ($m_{\rm v}$ = 12.7). Further spectra would help to constrain the spectroscopic and derived parameters. A more precise value of \vsini\ could also be found from an observation of the Rossiter-McLaughlin (RM) Effect \citep{Rossiter24, McLaughlin24, GW07}, which could provide an independent constraint of the age through gyrochronological relations \citep{Barnes07}. The slow rotation (\vsini\ = 2.4 $\pm$ 1.6 \kms) and faintness of the host star would make observations of the RM effect challenging but possible given the predicted amplitude of 35 ms$^{-1}$.

We find a small discrepancy between the stellar mass found from evolutionary models and an empirical calibrations based on eclipsing binaries, and take a mean of the two values in this analysis. WASP-37 is less massive than the Sun, \mstar\ = 0.925 $\pm$ 0.120 \msun, and is one of the older stars hosting a transiting planet at $\sim$11 Gyr. Although the age is relatively uncertain, it suggests that giant planet formation was taking place when the Milky Way was still relatively young. The model isochrone, shown in Figure \ref{evo}, suggests that WASP-37 is due to evolve from a main-sequence star into a subgiant at an age of $\sim$15 Gyr and subsequently move on to the giant branch after a further $\sim$1 Gyr. Although no meaningful constraint can be placed on the stellar tidal quality factor, $Q_{s}$, we note that if is greater than $10^{7}$  then the planetary spiral-in timescale \citep{Levrard09} is longer than the main sequence lifetime, and the planet will be engulfed by the expansion of the stellar radius during the giant phase less than 0.5 Gyr after the star has evolved onto the giant branch.

\begin{figure}[t]
\centering
\includegraphics[width=10cm]{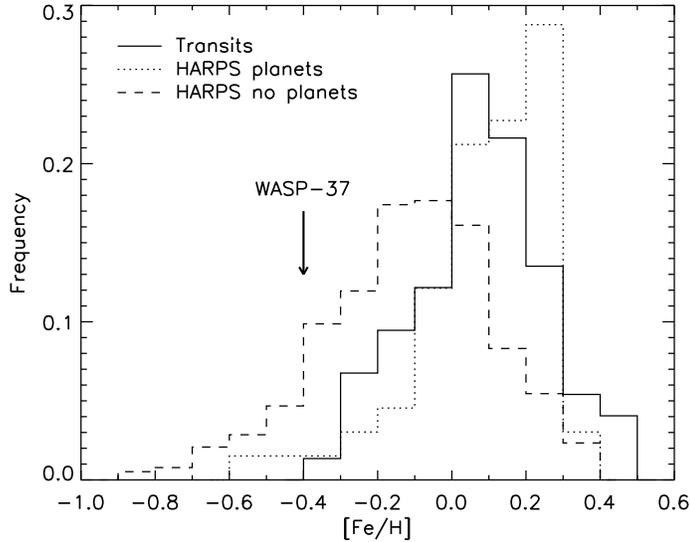}
\caption{Histogram of the metallicity of transiting planet host stars and stars with and without detections of RV planets from the HARPS GTO ``high-precision'' sample \citep{Sousa08}. WASP-37 is clearly in the tail of the metallicity distribution for stars hosting planets. }
\label{feh}
\end{figure}

WASP-37 has the second lowest metallicity ([Fe/H] = $-$0.4 $\pm$ 0.12) of the transiting planet host stars after WASP-21 \citep{Bouchy10}, and is firmly in the tail of the metallicity distribution for exoplanet host stars as shown in Figure \ref{feh}. The figure, similar to that in \citet{Ammler09}, shows the metallicity distributions for stars with transiting planets and stars with and without detections of RV planets from the HARPS GTO ``high-precision'' sample \citep{Sousa08}. The stars hosting transiting and RV planets show a very similar distribution, whereas the non-planet hosts appear to extend to much lower metallicities, suggesting that the heavy metal content plays a significant role in planet formation \citep{Gonzalez98, Santos04, FischerValenti05}.

\acknowledgments
The SuperWASP Consortium consists of astronomers primarily from QueenÕs University Belfast, St Andrews, Keele, Leicester, The Open University, Isaac Newton Group La Palma and Instituto de Astrofõsica de Canarias.  The SuperWASP-N camera is hosted by the Issac Newton Group on La Palma and WASP-South is hosted by SAAO. We are grateful for their support and assistance. Funding for WASP comes from consortium universities and from the UK's Science and Technology Facilities Council. FPK is grateful to AWE Aldermaston for the award of a William Penny Fellowship. Based on observations made at Observatoire de Haute Provence (CNRS), France and at the ESO La Silla Observatory (Chile) with the CORALIE Echelle spectrograph mounted on the Swiss telescope. The RISE instrument mounted in the Liverpool Telescope was designed and built with resources made available from QueenÕs University Belfast, Liverpool John Moores University and the University of Manchester. The Liverpool Telescope is operated on the island of La Palma by Liverpool John Moores University in the Spanish Observatorio del Roque de los Muchachos of the Instituto de Astrofisica de Canarias with financial support from the UK Science and Technology Facilities Council. The research leading to these results has received funding from the European Community's Seventh Framework Programme (FP7/2007-2013) under grant agreement number RG226604 (OPTICON). We thank Tom Marsh for the use of the ULTRACAM pipeline.

\bibliographystyle{aa}
\bibliography{bib.bib}

\end{document}